\documentclass[12pt]{article}

\usepackage{scicite}
\usepackage{subfig}
\usepackage{graphicx}
\usepackage{hyperref}
\usepackage{cleveref}

\usepackage{times}

\newenvironment{sciabstract}{%
\begin{quote} \bf}
{\end{quote}}

\newcounter{lastnote}

\title{Analysis of misinformation during the COVID-19 outbreak in China: cultural, social and political entanglements}

\author
{Yan Leng,$^{1\dagger}$ Yujia Zhai,$^{2 \; 3\dagger\ast}$ Shaojing Sun,$^{4}$ Yifei Wu,$^{5}$ Jordan Selzer,$^{6}$ \\
Sharon Strover,$^{7}$ Julia Fensel,$^{8}$ Alex Pentland,$^{1}$ Ying Ding,$^{9\ast}$
\\
\normalsize{$^{1}$Media Lab, Massachusetts Institute of Technology, Cambridge, USA} \\
\normalsize{$^{2}$Department of Information Resource Management,}\\
\normalsize{ Tianjin Normal University, Tianjin, China}\\
\normalsize{$^{3}$School of Information Management, Wuhan University, Wuhan, China} \\
\normalsize{$^{4}$School of Journalism, Fudan University, Shanghai, China}\\
\normalsize{$^{5}$Department of Computer Science and Technology, Tsinghua University, Beijing, China} \\
\normalsize{$^{6}$Department of Emergency Medicine, George Washington University, Washington, D.C., USA} \\
\normalsize{$^{7}$School of Journalism, the University of Texas at Austin, Austin, USA}\\
\normalsize{$^{8}$Westlake high school, Austin, USA}\\
\normalsize{$^{9}$School of Information, the University of Texas at Austin, Austin, USA}\\
\normalsize{$^\ast$To whom correspondence should be addressed;}\\
\normalsize{E-mail: ying.ding@ischool.utexas.edu, yjzhai@tjnu.edu.cn}\\
\normalsize{$\dagger$Equal contribution.}
}

\date{}

\begin{document} 


\baselineskip24pt


\maketitle


\begin{sciabstract}
COVID-19 resulted in an infodemic, which could erode public trust, impede virus containment, and outlive the pandemic itself. The evolving and fragmented media landscape is a key driver of the spread of misinformation. Using misinformation identified by the fact-checking platform by Tencent and posts on Weibo,  our results showed that the evolution of misinformation follows an issue-attention cycle, pertaining to topics such as city lockdown, cures, and preventions, and school reopening. Sources of authority weigh in on these topics, but their influence is complicated by peoples' pre-existing beliefs and cultural practices. Finally, social media has a complicated relationship with established or legacy media systems. Sometimes they reinforce each other, but in general, social media may have a topic cycle of its own making. Our findings shed light on the distinct characteristics of misinformation during the COVID-19 and offer insights into combating misinformation in China and across the world at large. 
\end{sciabstract}

\section*{Introduction}
While the world is struggling to fight the COVID-19 pandemic, another more insidious epidemic—an infodemic, the term coined by WHO Director-General Tedros Adhanom Ghebreyesus to describe the prevalence of misinformation on the coronavirus—is threatening the health and safety of the public.
Misinformation has existed throughout history; however, with the rise of social media platforms—Twitter, Facebook, Reddit, Weibo, microblogs, among others—its spread has transcended borders and significantly increased in both pace and audience~\cite{zarocostas2020fight}.  Understanding the flow and spread of misinformation is a vital issue to reduce its negative impacts.  Just as COVID-19 has now spread to millions worldwide, misinformation, ranging from pseudoscience to conspiracy, has also spread online at an alarming rate, stirring panic and causing confusion. In China, such panic and confusion have led to massive scrambling for fictitious "cures," such as garlic and Shuanghuanglian, a traditional Chinese medicine. 
Misinformation is a complex and elusive concept to rigorously define, despite the growing volume and scope of research on this topic. Briefly, it refers to "information considered incorrect based on the best available evidence from relevant experts at the time "~\cite{vraga2020defining}. In the past decade, the explosion of social media has significantly increased the dissemination of misinformation pertaining to numerous topics, including health, politics, and entertainment. As a result, identifying and combating health-related misinformation on social media has become a top concern of public health~\cite{chou2018addressing}. During the 2018-19 Ebola outbreak, more than 25 percent of people in DR Congo believed misinformation about the outbreak, which significantly impeded public adoption of preventive behavior~\cite{vinck2019institutional}. Factors affecting one's response to misinformation are complex and varied. Social norms, as well as the compatibility between the misinformation and one's belief system, have a significant influence on one's acceptance of misinformation~\cite{del2016spreading}.  
To date, the majority of research on misinformation has examined Western media platforms; however, Chinese social media constitutes a very significant market. Weibo hosts more than 556 million registered users, with 313 million considered active monthly, compared to 319 million on Twitter~\cite{huang2017}.
There are some differences when comparing Weibo and Twitter, however. Past research has found that Weibo demonstrates a different pattern of information dissemination than Twitter, though the two platforms are often viewed as comparable~\cite{chen2012follow}. Twitter is more open to contestation and plural views, whereas Weibo is under stricter control and draconian censorship by the Chinese government~\cite{king2017chinese}.  Besides, compared to Twitter, the network connections on Weibo are more hierarchical because users tend to follow others at a similar or higher social level~\cite{huang2017}. 
Considering the consequential impact of misinformation,  it is vital to investigate the spread of misinformation in the ongoing COVID-19 pandemic. China—the first country undergoing the massive infection and lockdown—provides a rich cultural context in which to investigate COVID-19 related misinformation.

\subsection*{Event Driven Misinformation}
Using the data collected from the Jiao Zhen platform from January 19, 2020, to March 29, 2020, we examine how misinformation (rumors and pseudoscience) and gossip (inconclusive information) evolved throughout the different stages of COVID-19 epidemic in China, and how its dissemination correlates to significant news events and government policies. It should be noted that the lockdown was around the time of the Spring Festival holiday--the most celebrated tradition in China. During this time, local residents and migrant workers return home for family reunions, generating the most significant human migration on the planet.  As such, it is also the time when the government is on high alert to ensure social order. The approaching of the lunar-year festivity served to further amplify the abruptness of the lockdown measure, feeding public uncertainties about the seriousness of the virus. 

\begin{figure*}[tbhp]
\centering
\subfloat[Temporal distribution]{\includegraphics[width=.7\linewidth]{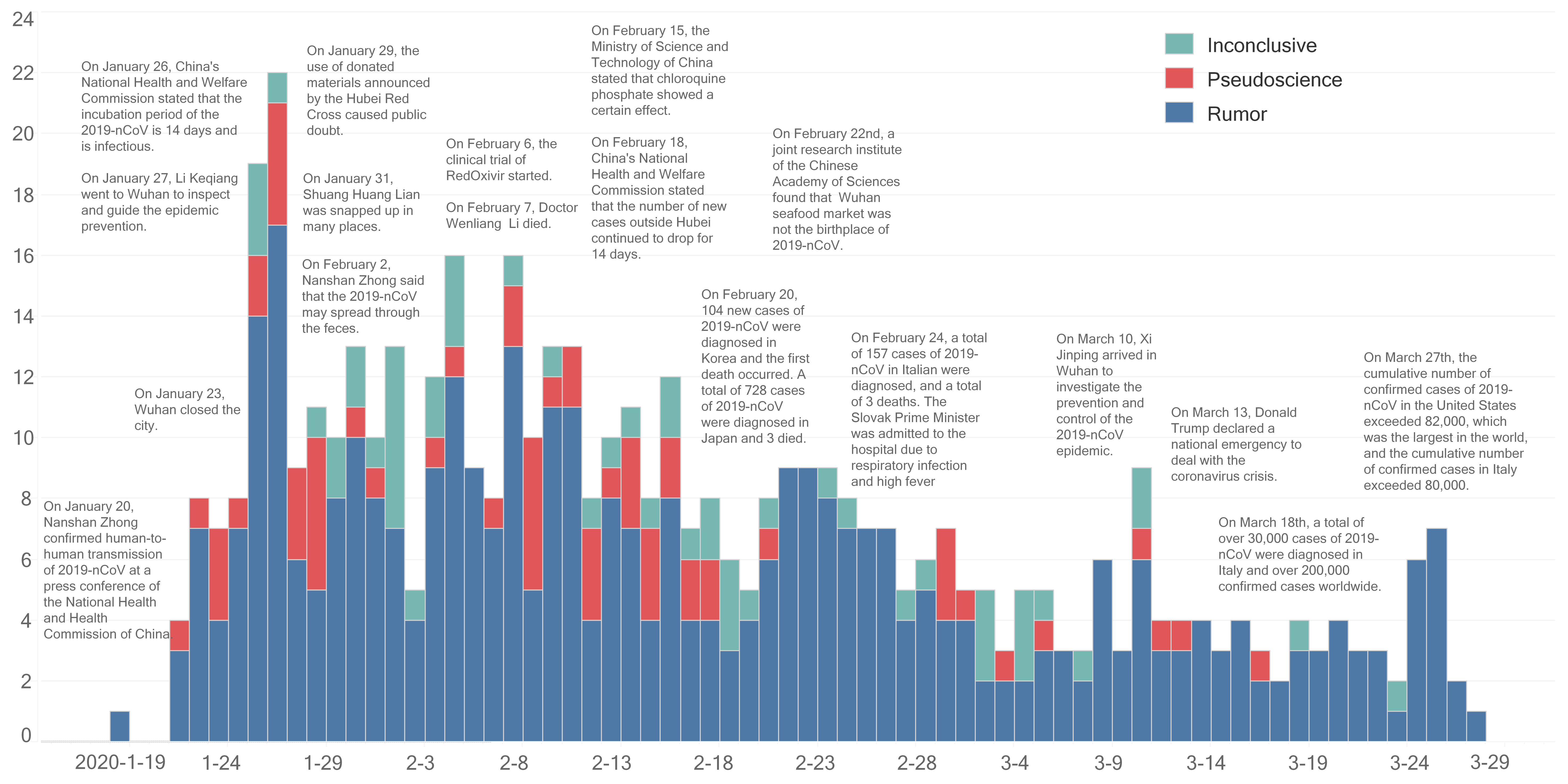}\label{fig:temporal_misinfo}}\\
\subfloat[before February, 2020]{\includegraphics[width=.25\linewidth]{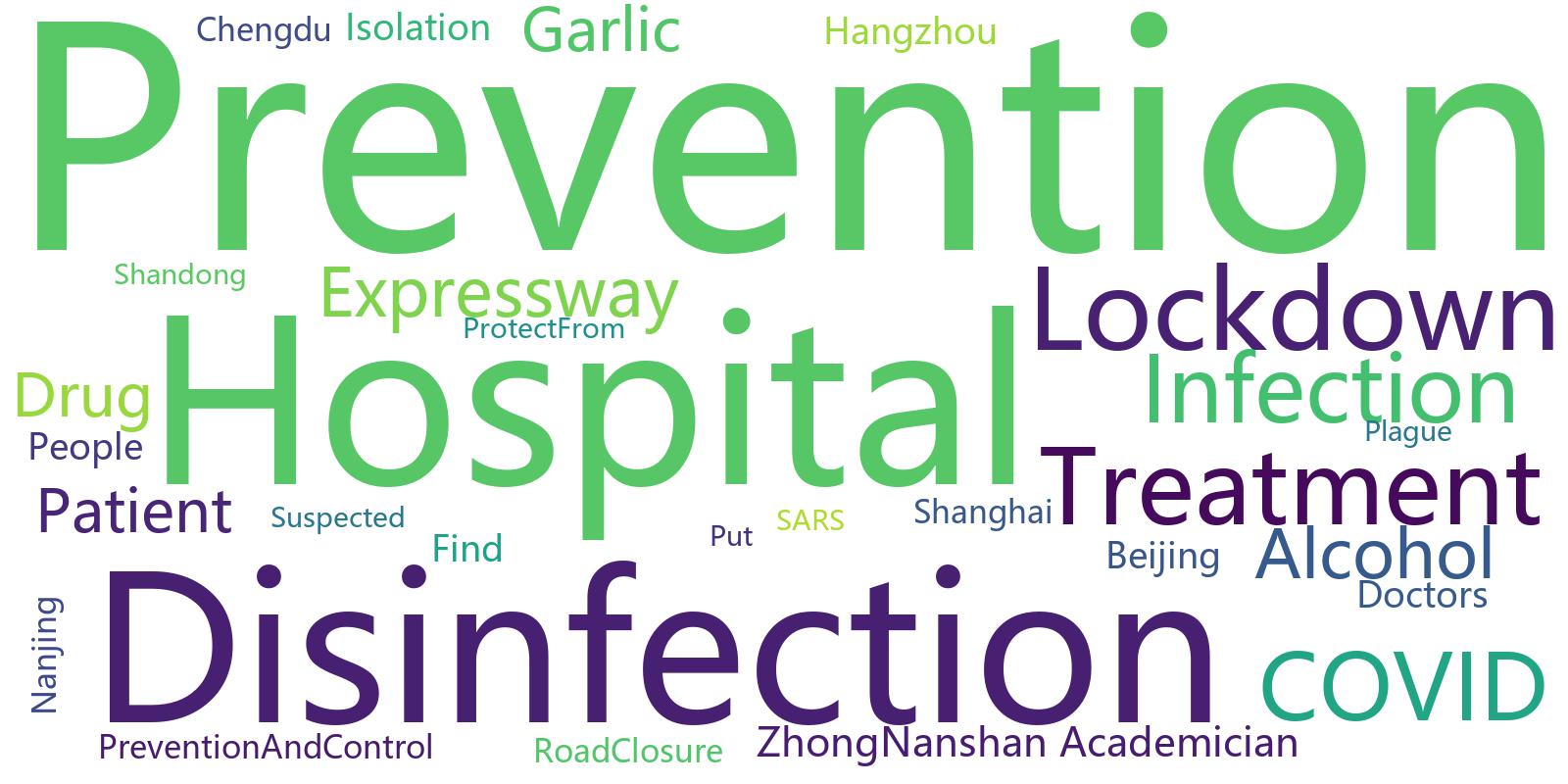} \label{fig:wc1}}
\subfloat[February, 2020]{\includegraphics[width=.25\linewidth]{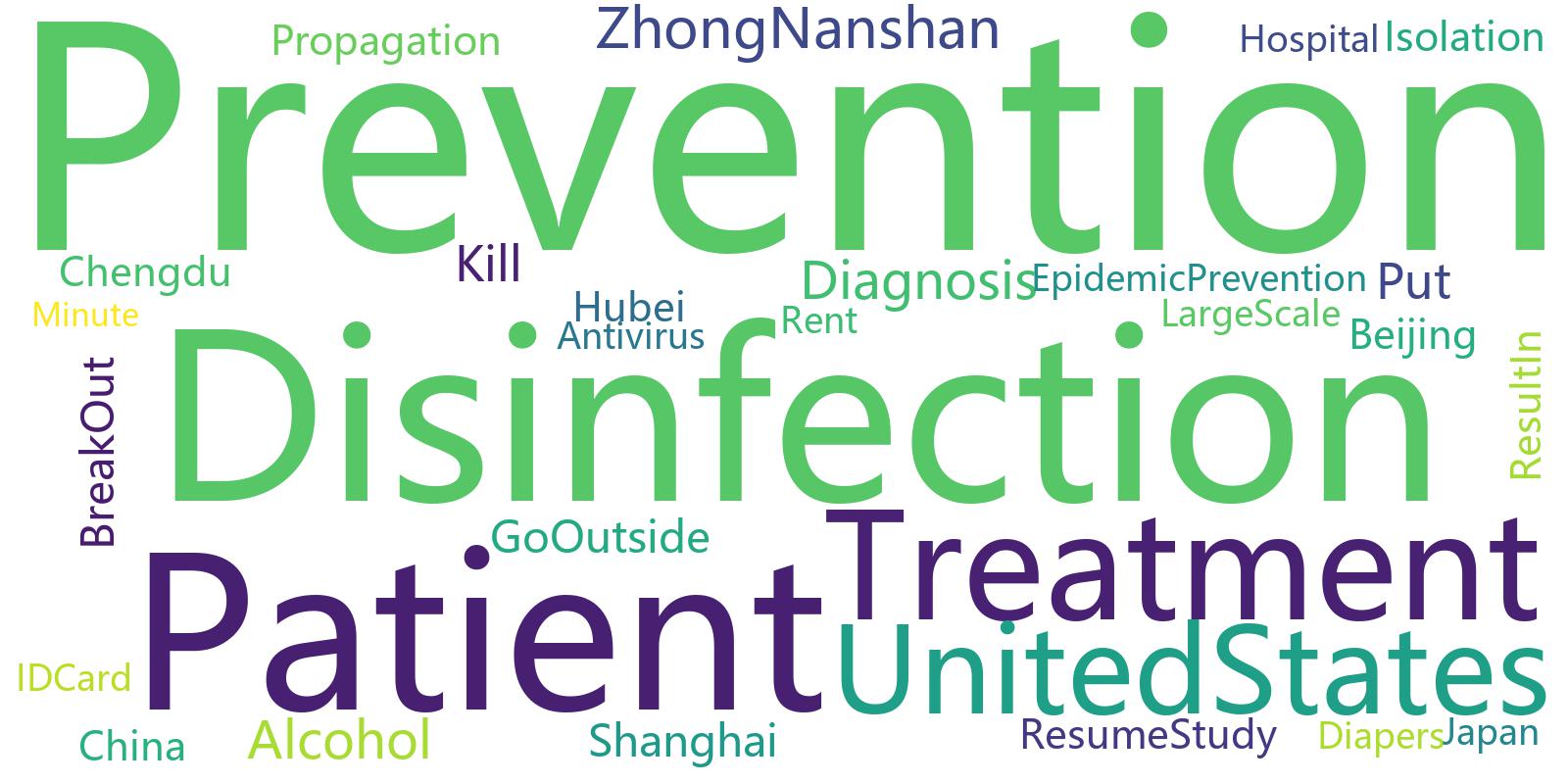}\label{fig:wc2}}
\subfloat[March,  2020]{\includegraphics[width=.25\linewidth]{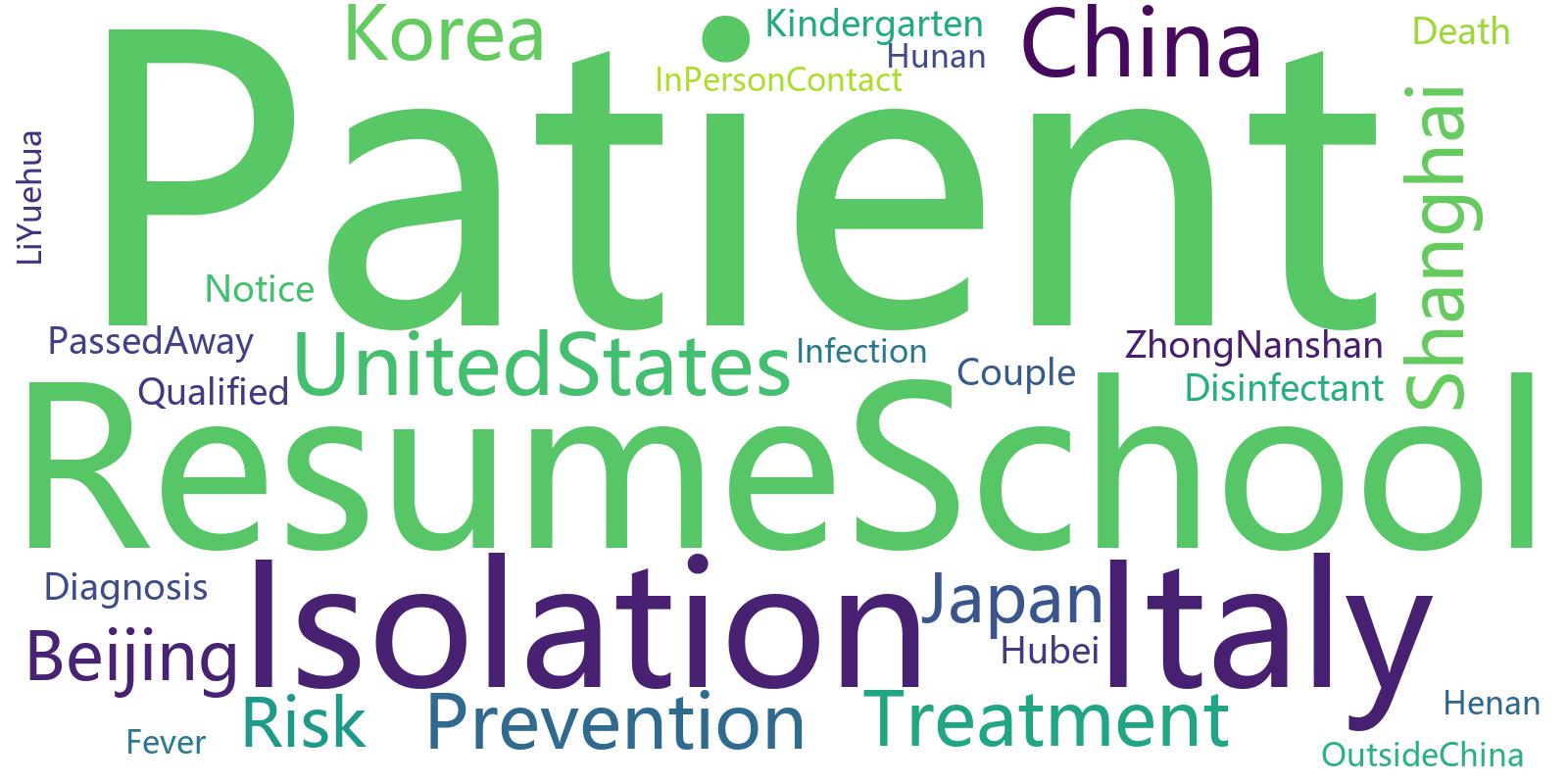}\label{fig:wc3}} \\
\subfloat[Summary of sensationalist misinformation]{\includegraphics[width=.7\linewidth]{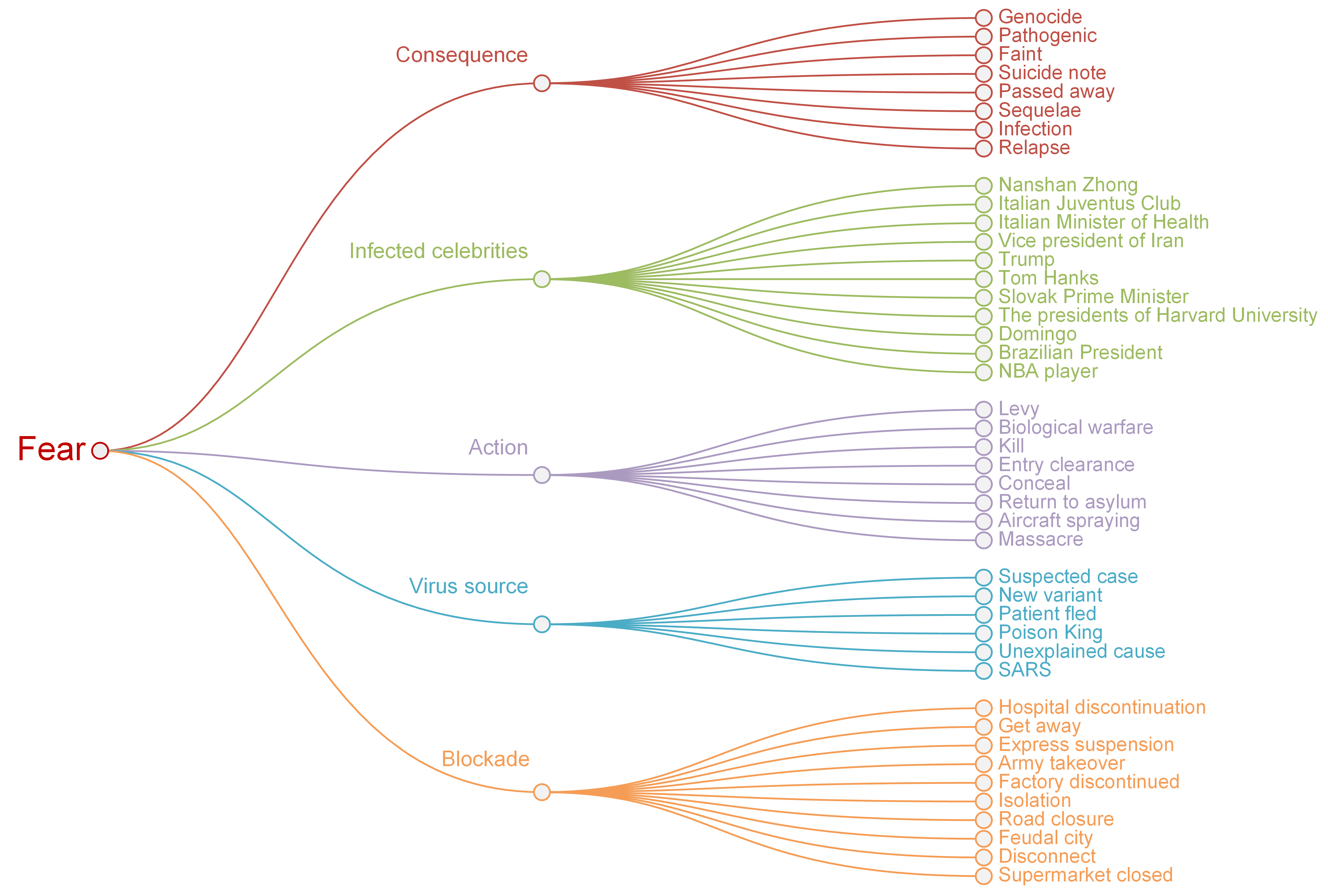}\label{fig:fear}}
\caption{{\bf Temporal variations of misinformation.} (a) Count of misinformation. The x-axis corresponds to the date and the y-axis is the number of posts for different types of misinformation, with the corresponding category shown in the legend. We label major events (e.g., policies or scientific news) with vertical lines. (b)-(d) Popular topics of misinformation. The size of the word is proportional to the frequencies of occurrences from the titles of misinformation. For clearer comparison, we removed some high frequency words that were popular throughout the whole period, including novel coronavirus, infection, facial masks, Wuhan, epidemics.
(e) is a categorization over the misinformation that may cause fear. }
\label{fig:main_1}
\end{figure*}

In Figure~\ref{fig:temporal_misinfo}, we present the temporal trend of misinformation and gossip, which covers three waves of misinformation related to city lockdowns, transmission and cures, and reopening schools. The first and most prominent peak happened shortly after the lockdown was announced in Wuhan, which occurred on January 23, 2020. 
In response to this unprecedented policy, there was a surge of sensationalizing information reporting severe disruptions in Wuhan (e.g., gas stations within Wuhan had stopped operating) and similar policies in other cities in China (e.g., a lockdown in Hankou in Wuhan and Xinyang in Henan). 
On February 2, Nanshan Zhong, a well-respected pulmonologist and advisor to the Chinese government, announced that COVID-19 might spread via fecal transmission~\cite{nanshanzhong_feikou}. After this, a spike of misinformation regarding possible modes of transmission, cures, and prevention methods occurred. For example, a number of false claims that the virus was spread through mosquitoes, flies, and through the sewer were posted. 
A third peak occurs around March 9, when the epidemic transitioned into a containment stage.  During this peak, we observe a transition of misinformation related to lifting the lockdowns, restarting schools, and news in foreign countries (e.g., Italy and the U.S.).

{To understand the evolving topics, we present the word cloud, split by calendar month in Figure~\ref{fig:wc1}--~\ref{fig:wc3}. 
Figure~\ref{fig:wc1} covers the topics in January. The five most common words are  {\it prevention, disinfection, hospital, lockdown, and treatment.}. 
A unique aspect of misinformation during this period is the manipulation of the original lockdown in Wuhan into lockdowns of other cities and ridiculous policies in Wuhan.

In February (as in Figure~\ref{fig:wc2}),
{\it prevention, disinfection, treatment} remained popular within misinformation. Meanwhile, new topics emerged, such as patients, the U.S., and  Nanshan Zhong. 
As the epidemic spread further and the testing capacity increased, we observe more sensationalist misinformation about patients, e.g., an outbreak in a specific city, or a confirmed patient who contacted a large number of people. 
In the latter half of February, the U.S. attracted a lot of attention in the media in China; most of the misinformation was related to the U.S.'sU.S.'s confirmed cases and potential for an outbreak. 
Meanwhile, we also observed manipulations of Nanshan Zhong's statement.  Misrepresenting or manipulating an authoritative figure, is a well-known misinformation strategy~\cite{fragale2004evolving}.  Linking a credible/authoritative source to misinformation tends to amplify the believability of the untruth. For example, a piece of misinformation said that "Nanshan Zhong's team found that the infection rate of COVID is higher on smokers than non-smokers."
This is the intentional exploitation of a reputable source to gain legitimacy. Unfortunately, however, it confuses and serves to undermine the credibility of the true experts and authoritative institutions.  This example highlights the importance of both the audience and social platforms of vetting the credibility of a claim by confirming the source of the news.


In March, when the epidemic transitioned into a containment stage, {\it resume school, isolation, Italy, China, US} became the focus, as shown in Figure~\ref{fig:wc3}. 
There was a noticeable shift in the focus compared with January and February with less misinformation about prevention, disinfection, and treatment. 
Interestingly, there is a growing focus in resume school and Italy.  
}

Throughout COVID-19, one prominent characteristic of misinformation is sensationalism and scaremongering. Sources of misinformation manipulate true stories, forage ridiculous news, and unverified sources to appeal to the crowds emotionally. Sensationalism in the media has long been a popular topic of fierce debate. Despite the absence of an authoritative definition of media sensationalism, researchers tend to agree that sensational messages are likely to provoke senses and emotions. Furthermore, both content features (e.g., a topic about crime typically is more sensational than other topics) and formal elements (e.g., using a colorful font is more sensational than plain font) have proven to contribute to sensationalism (Grabe et al., 2001). From a linguistic angle, Molek-Kozakowska (2013) identified a range of pragmatic devices that may arouse audiences' sensational feelings. Dubbed as sensational illocutions, the methods include exposing (e.g., denouncing a crime), speculating (suggesting future consequences), generalizing (extrapolating to a larger population), warning (generating anxiety) and extolling (extraordinary facts/events/individuals). Integrating the perspectives mentioned above on content, form, and pragmatics about sensationalism, the researchers followed the typical procedure for content analysis. They classified the misinformation into five categories, with a final intercoder agreement of .94. In Figure~\ref{fig:fear}, the categories of misinformation comprise exposing virus (e.d., the origin of the virus), extraordinary incidence (e.g., infected celebrity), warning of risk severity (e.g., relapse), conflict/conspiracy (e.g., biological warfare), speculating hard life (e.g., lockdowns). Take the category of exposing viruses as an example: several posts mentioned unknown virus sources, stories of diagnosed patients fleeing quarantine, or claims that one patient infected a large number of others. Anxiety and emotions may also be entangled with false claims of lockdowns of different cities, expressways, and even at gas stations.

\subsection*{Collective sensemaking and profiteering: tracing  misinformation and gossip on cures and prevention}
Under periods of such uncertainty and anxiety, the public is more susceptible to misinformation, which in turn self perpetuates. Some research suggests that social media users, in general, may be more vulnerable to misinformation as well~\cite{burki2019vaccine}. 
As a major threat to public health, COVID-19 has triggered a surge of misinformation on cures and prevention.  
This information can be harmful for numerous reasons. 
Some of the misinformation may contribute to direct physical harm; for instance, claims that firecrackers and drinking boiling water will cure the virus.  

\begin{figure*}[tbhp]
\centering
\subfloat[Evolution of the misinformation posts on cures and prevention. 
]{\includegraphics[width = 1\linewidth]{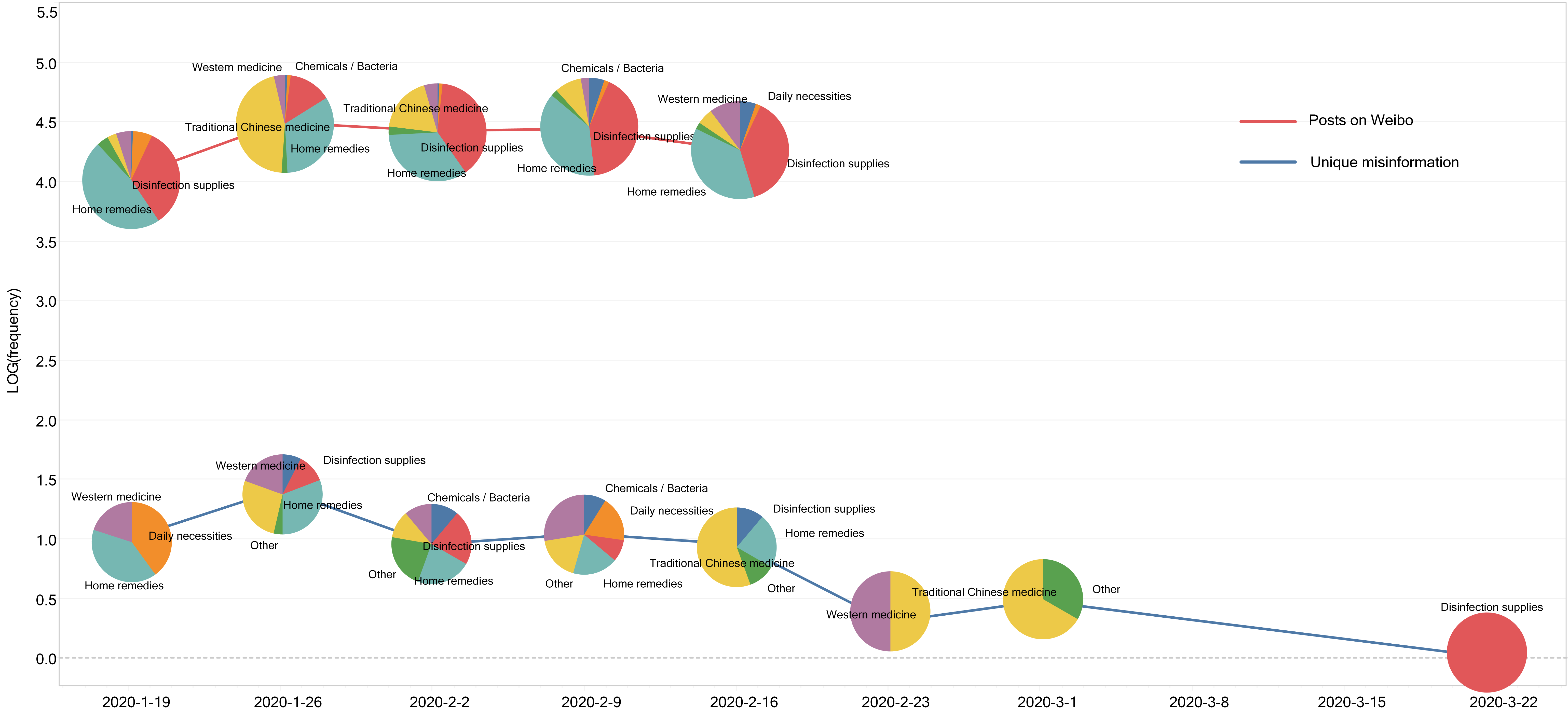}\label{fig:misinfo_dynamics}}\\
\subfloat[Temporal popularity of the top 10 misinformation posts on cures]{\includegraphics[width = 1\linewidth]{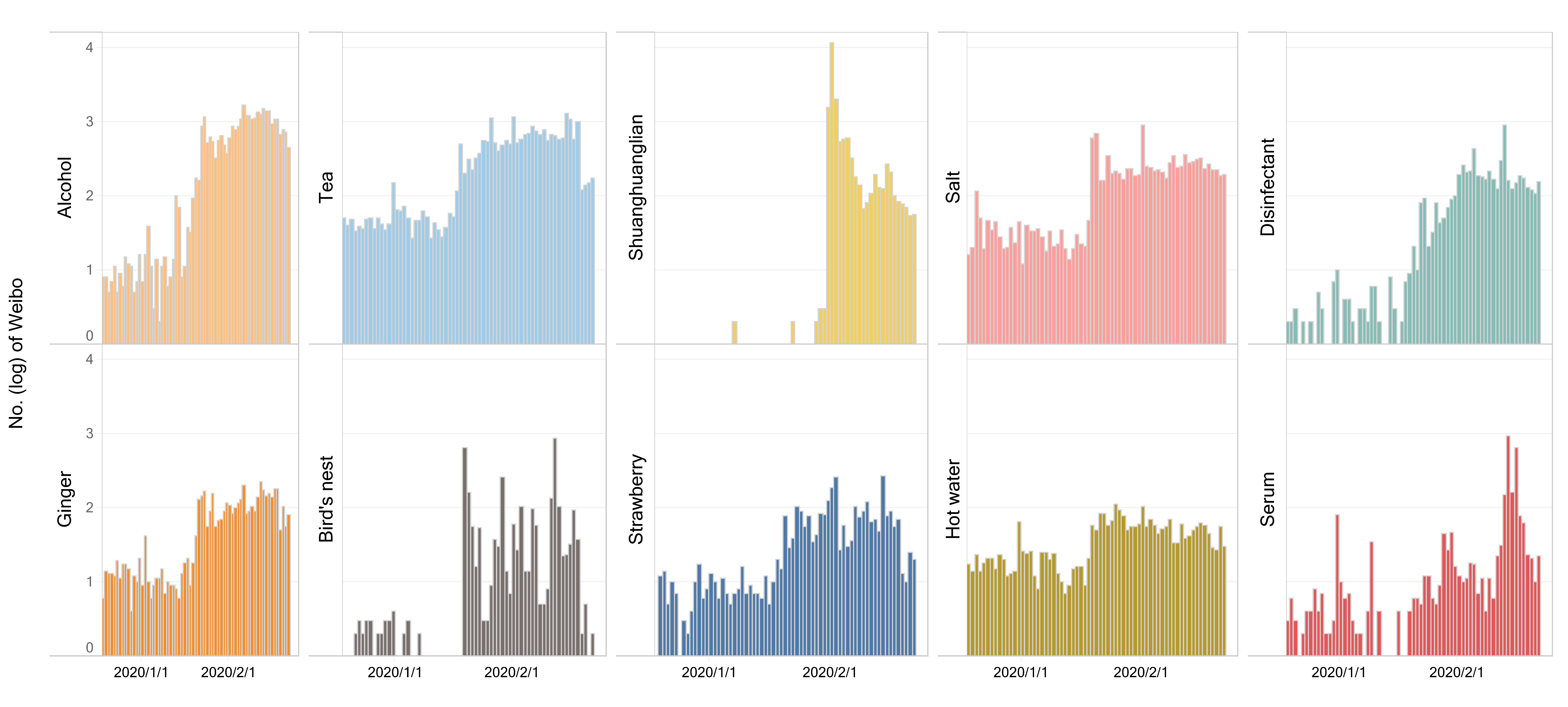} \label{fig:top10_cures}}
\caption{{\bf Evolution of misinformation regarding cures and prevention for COVID-19.} We categorize different types of cures and prevention into food, Chinese medicine, Western medicine, disinfection suppliers, daily necessities, and chemicals/germs. In the early stage of the epidemic, there was a large amount of misinformation mentioning home remedies. As time progressed, there was a shift towards Chinese medicine, especially in late January to early February, with Shuanghuanglian as a representative example. The patriotism and cultural preservation advanced by the Chinese government added to the ideas about the effectiveness of Chinese herb medicine for COVID-19.  Throughout the epidemic there was constant misinformation regarding Western medicine; some were conventional medicines (e.g., aspirin and mucosolvan), and others were more recently approved or studied (e.g., oseltamivir and resveratrol).  Interestingly, as the epidemic in China shifted into the containment stage in mid-March, the misinformation regarding cures and prevention almost completely disappeared. }
\label{fig:treatment}
\end{figure*}

In Figure~\ref{fig:treatment}, we analyzed the evolution of misinformation on cures and prevention. We observe a large number of rumors regarding home remedies, at the beginning of the COVID-19 outbreak.
Among the posts on Weibo about cures and prevention, home remedies attracted the most extensive attention, followed by disinfection supplies and traditional Chinese medicine, as shown on the upper panel of Figure~\ref{fig:misinfo_dynamics}. 
Specifically, the interests in home remedies took up more than 36.4\%  among all posts on misinformation throughout the observational period. Figure~\ref{fig:top10_cures} presents the top 10 cures and prevention methods mentioned, based on the number of related posts in the Weibo. There is a constant interest in alcohol, tea, salt, disinfection supplies, and ginger. We also observed a significant spike in Shuanghuanglian.

There are a variety of reasons to explain the popularity of home remedies in this context. 
First, the public's high familiarity with home remedies, connecting closely with day-to-day life experience, could result in intuitive thinking as compared to critical or cognitive thinking~\cite{effron2019misinformation}.
Many of these home remedies, rooted in Chinese culture, have been on the market for quite some time (e.g., vinegar, alcohol, and garlic). 
Some people utilized past experience and unconfirmed knowledge on home remedies to come up with potential solutions, which might be part of collective sensemaking to resolve the anxiety in response to the ongoing health crisis~\cite{StarbirdMisin}. Moreover, with repeat encounters in the past, the crowd is more likely to trust and endorse the widespread, though unproven, use of home remedies~\cite{berinsky2017rumors}, which also could explain why some untested home remedies become constants in  misinformation~\cite{tencent}. 
Second, pandemic profiteering and price gouging involving face masks and fictitious cures emerge instantaneously.  Some unscrupulous actors take advantage of this as a source of revenue generation to sell their home remedies by making false or unsubstantiated claims. 
For example, individuals attempt to gain profits by selling mooncakes using SHL as an ingredient. 
Additionally, as home remedies become popular, some Weibo users seize this opportunity to attract more followers in an attention driven-economy~\cite{simon1996designing}.


\subsection*{Fighting the infodemics: Shuanghuanglian as a case study} 
\begin{figure*}[tbhp]
\centering
\subfloat[Cascade of Weibo]{\includegraphics[height = .3\linewidth]{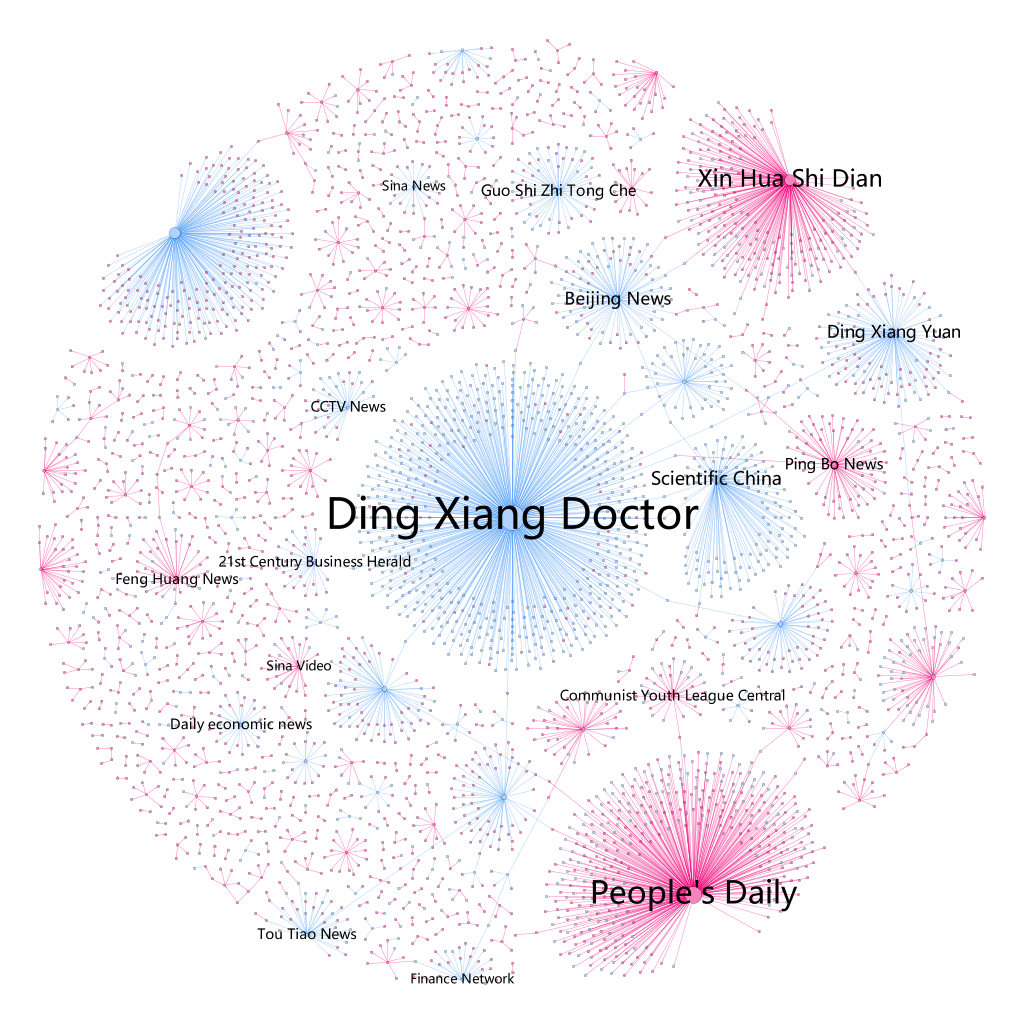} 
\label{fig:shl_diffusion}}
\subfloat[Popular tags in Weibo ]{\includegraphics[height = .3\linewidth]{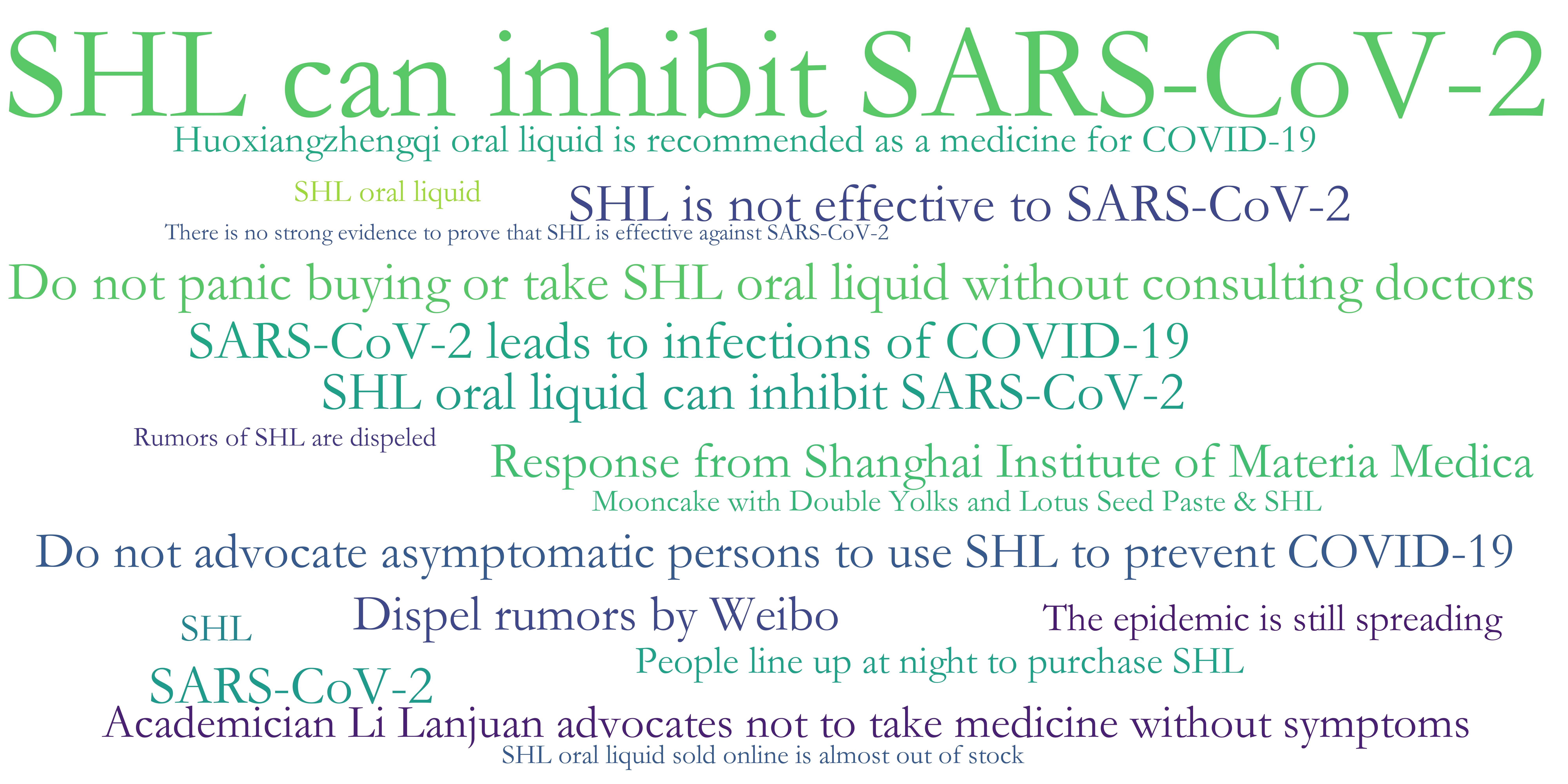} \label{fig:shl_tags}}
\caption{{\bf SHL-related Weibo.} (a) Cascade of Weibo related to SHL. Users who are against the information (e.g., posted "fake, false, should not use, no, negative") are colored in blue. The remaining are colored in red. We removed users who are not connected to any other individuals. Accounts with a large number of posts are labeled by their account name. 
On January 31, 2020, Xin Hua Shi Dian, a party-owned publisher, cited this article and quoted: "SIMM and Wuhan Institute of Virology had discovered that SHL herbal remedy could "inhibit" 2019-nCov",  and mentioned that SHL is currently under clinical trial in hospitals in Shanghai and Wuhan. Nine minutes later, People's Daily, with 1.1 billion followers, endorsed the same news. 
Four hours later, Ding Xiang Doctor, an online community for physicians, raised the concern that such news will cause a shortage, and therefore people should \textbf{not} use it for prevention. 
Four hours later, a personal account with 2.7 million followers, shared an ironic post, which is surprisingly more accessible than many other accounts with more followers (e.g., Li Shi Pin, which has nine times more followers). 
This account held negative opinions about Chinese medicine and specifically satirized about a person's daily life with seven types of Chinese herb medicine. 
Six hours later, Scientific China, a science account with 3.5 million followers, shared the post with a catchy title, asking whether SHL can inhibit novel Coronavirus. 
SHL soon went viral on Weibo. 
(b) Popular tags in Weibo related to SHL. The size of the tag is proportional to the number of mentions. }
\end{figure*}

In order to reduce harm to the public, Director-General Dr. Tedros Adhanom Ghebreyesus claims, "it is essential to fight infodemics with information "~\cite{the2020covid}. However, information itself is not enough to accomplish this goal. 
Due to a lack of comprehensive scientific training of the public and even some news media, misunderstandings or inaccurate interpretations of scientific studies may intensify the fear and panic. 
As an example, we investigate the claim that Shuanghuanglian (SHL) can {inhibit} Coronavirus.
On January 31, 2020, the Shanghai Institute of Materia Medica (SIMM, under the Chinese Academy of Sciences) and the Wuhan Institute of Virology had discovered that SHL herbal remedy could "inhibit" 2019-nCov~\cite{shl_article}.
The study was based on a laboratory in vitro studies and required further clinical studies to confirm its effectiveness on humans. 
However, the crowd interpreted the research as 'SHL helps to prevent or cure coronavirus.' 
Misinterpreting the result in an environment of significant uncertainty contributed to a wave of panic-buying of SHL.  
The diffusion of SHL is shown in Figure~\ref{fig:shl_diffusion}, which is especially interesting due to the polarized news and discussions, as revealed by the tags from the posts in Figure~\ref{fig:shl_tags}. 
Specifically, the discussion revolves around the effectiveness and ineffectiveness of SHL. 
In response to the panic-buying of the crowd, there is a caveat against such behavior. 

Individuals are more likely to believe information with a logical narrative from a source they perceive to be credible~\cite{scheufele2019science}. Source credibility varies, and may even be related to what sort of media people use.  A post that SHL was effective for \textit{inhibition} of COVID-19 was endorsed by two credible news media outlets, both providing an abbreviated scientific explanation. However, due to a lack of understanding of science and the critical omission of a clear explanation by the media, many understood the report as claiming SHL to be an effective \textit{treatment} for COVID-19.
This highlights the importance of the press to be extremely accurate and specific about scientific facts that are presented.~\cite{leshner2016communicating}. 
A similar phenomenon was observed in the U.S. with hydroxychloroquine, causing a shortage of the medication for lupus and rheumatoid arthritis patients, after repeated claims of its (unproven) efficacy to treat COVID-19 by President Donald Trump~\cite{jakhar2020potential}.

It is known that ineffective scientific communication can be costly to society~\cite{fischhoff2013sciences}. This case study highlights the extraordinary significance of scientific communication in pandemics. 
For scientists,  understanding the significance of adequately communicated findings is paramount. 
It is especially important for the scientific media to interpret the evidence, and to correct the general media when it has misleading coverage.
Press and media must understand how to serve as the bridge and even translator between the scientists and the general public. 
Lastly, the public must be critical about the direct interpretation of scientific reports without adequately understanding and searching for reputable coverage and information before making decisions. 

\section*{Discussion}

Our findings on the evolution of misinformation speak to how the attention cycle--a concept describing the complicated relationship between issue development and corresponding attention from media and the public - operates during an epidemic. 
Attention cycles interact with the ways that journalists frame issues, but social media communication patterns have independent trajectories that build on the emotions that the issue and the platforms may encourage~\cite{shih2008media}
(Jordon, 2001; Shih et al., 2008).

Henry and Gordon (2001) argued that the public tends to pay more attention to an event when it is novel, and such attention will shift to other happenings or new aspects of the event as people become familiar with or bored of the incident~\cite{henry2001tracking}. The issue cycle is related to the influence of various factors such as journalistic practice, cultural values, and media systems, and how individuals choose to use media and evaluate credible messages. The present study showed that discussion of misinformation follows a unique issue cycle. 

Attention cycles interact with the ways that journalists frame issues. Still, social media communication patterns have independent trajectories that build on the emotions that the issue and the platforms may encourage.

Attention cycles interact with the ways that journalists frame issues, but social media communication patterns have unique trajectories that build on the emotions that the issue and the platforms may encourage~\cite{henry2001tracking, shih2008media}. 
The example of how  Chinese medicine and Western medicine play into unfounded beliefs illustrates the ways social and cultural belief systems clash with scientific sources. Past research suggests that social media and traditional media interact with each other, affecting public attention/discussion of an issue and leading to inter-media agenda-setting. 

Our findings do show that traditional mainstream media--such as Xinhua News Agency and People's Daily along with other popular market-driven or personal media accounts--play a role in the arena of misinformation. Such finding echos the call for investigating the changing media landscape through a holistic and inter-media perspective.  
Our study also offers insight into combating misinformation in a massive public health crisis. As Lewandowsky et al. commented, debunking misinformation merely from a science communication perspective may not be sufficient to mitigate the adverse effects~\cite{lewandowsky2017beyond}. Instead, incorporating cultural, political, and social ingredients into the process of protecting the public from the threat of misinformation can yield superior outcomes. In a time of an infectious disease crisis, people are more likely to believe in unproven cures such as Shuanghuanglian. Social media amplifies emotional messages and responses; the platforms themselves contribute to fast and unthinking responses, thus increasing the possibilities that misinformation will circulate widely.  Seemingly naive beliefs, such as the efficacy of traditional herbal medicine, even without rigorous scientific evidence, may give residents feelings of security and control. 
Sometimes, such misinformation about unproven cures can be perpetuated by governments and other stakeholders. As noted earlier, Donald Trump has repeatedly recommended the use of hydroxychloroquine--a yet unproven treatment for COVID-19 which has well known potential risks. Meanwhile, the cacophony of different stakeholders may crowd out the authoritative opinions of legitimate experts. Therefore, only when the public, politicians, media, and other stakeholders are well informed and genuinely concerned with the public good, can they together push public discourses closer to the truth.  
The Shuanghuanglian case also speaks to the impact that a lack of consensus in the science community may have on the general public.  When multiple sources report conflicting information, it can be difficult for the layperson to sort out truth from fiction. 
When and how to convey expert disputes to the public is a practical and meaningful question. Notably, during a pandemic crisis, the timing and amount of publicizing information about expert disputes warrant more attention and research. 
Finally, our results also raise questions about the responsibilities of the different components in the media ecosystem in responding to infodemics.  People using social media, the scientific community is attempting to provide accurate and timely information, governments attempting to productively intervene in a crisis, and platforms that carry and algorithmically manipulate the messages all have essential roles in combating disasters, rumors, and damaging effects of misinformation.   
It would be valuable for future research to investigate how any potential patterns or cycles of misinformation vary across disasters, especially pandemics of emerging infectious diseases. Wu et al. (2016), analyzing Ebola-related rumors on microblogs in China, found that rumors tend to be related to real events associated with the virus. Specifically, rumors seem to concentrate on the source of the virus, fear about the virus, preventive measures, among others. Such a finding is, in general, consistent with that of the present study. Considering the nature of emerging infectious diseases, representative topics--such as a source of the pathogen, detection, and prevention-- are more likely to generate misinformation. Future research should also look at how misinformation about pandemics differs from other crises such as earthquakes and political unrest. More concretely, it would be enlightening to compare types of rumors across different disasters or infodemics. Is there a typical lifecycle to what people attend to at different phases of a crisis? Also, comparative research across countries would be valuable too. Consider what is going on in the U.S. where a seeming lockdown is going on, but it is more voluntary (monetary fines being a penalty, for example) than in China. Such a comparison reflects the role of the State, extending into more important questions about democracy and the struggle with individual rights v. collective good. Research along this line will inform worldwide joint efforts to combat a pandemic like Covid-19 too.

\paragraph{Data} We collected Weibo posts identified as misinformation or gossip from Tencent, the Jiao Zhen Platform~\footnote{This platform can be accessed through \href{https://vp.fact.qq.com/home}{this link}}, from 01/18/2020 to 03/27/2020. There were 531 COVID-related posts identified as misinformation or gossip in total. 
The Jiao Zhen Platform is a crowdsourcing platform where users can dispel posts on Weibo as rumor or identify as inaccurate. The platform requests identification of credentials, such as education and affiliated institution, as well as an explanation by  either citing the original source of information or a research article proving it to be false. 
The platform employs a full staff to verify the information once a record has been uploaded~\footnote{More information regarding the platform can be accessed through the following \href{https://cloud.tencent.com/developer/news/17550}{link} in Chinese. }.

We also collected 3.3 million Weibo posts from January 18, 2020, to February 26, 2020, using the advanced search function of the Weibo keyword indexing. The keywords we used included COVID-19, novel coronavirus, corona, epidemics, novel pneumonia, pneumonia in Wuhan.
We collected the data on February 26, 2020. 
We started from midnight on January 18, 2020, and each query searched all the posts within an hour. 
Each time, 50 pages were returned, each contained around 20 posts. 
Therefore, if the number of posts within that hour exceeded 1000, the information could no be collected in full due to the limitations of the search function. 
We continued this process control until the end of February 26, 2020.

\bibliographystyle{Science}

\bibliography{scifile}



\clearpage


\end{document}